\documentclass[aps,showpacs,nofootinbib,superscriptaddress]{revtex4}
\usepackage{epsf}
\usepackage{graphicx}
\usepackage{amsmath}
\def\slashchar#1{\setbox0=\hbox{$#1$}
   \dimen0=\wd0 \setbox1=\hbox{/} \dimen1=\wd1
   \ifdim\dimen0>\dimen1 \rlap{\hbox to \dimen0{\hfil/\hfil}} #1
   \else  \rlap{\hbox to \dimen1{\hfil$#1$\hfil}} / \fi}

\begin{document}

\title{$\chi$-SU(3) BETHE SALPETER MODEL: EXTENSION TO SU(6) AND SU(8)
SPIN-FLAVOR SYMMETRIES}

\author{J. NIEVES, C. GARC\'IA-RECIO, L.L. SALCEDO }

\affiliation{ Departamento de
 F\'\i sica At\'omica, Molecular y Nuclear, Universidad de Granada \\
 E-18071 Granada, Spain}

\author{V. MAGAS, A. RAMOS}
\affiliation{Departament d'Esctructura i
  Constituents de la Mat\`eria, Universitat de Barcelona\\ 
E-08028 Barcelona, Spain}

\author{T. MIZUTANI}
\affiliation{Departament of Physics, Virginia
  Polytechnic Institute and State University\\ Blacksburg, VA 24061,
  USA}

\author {H. TOKI}
\affiliation{Research Center for Nuclear Physics (RCNP), Osaka University \\
Ibaraki, Osaka 567-0047, Japan}

\begin{abstract}
Consistent SU(6) and SU(8) spin-flavor extensions of the SU(3) flavor
Weinberg-Tomozawa (WT) meson-baryon chiral Lagrangian are constructed,
which incorporate vector meson degrees of freedom.  In the charmless
sector, the on-shell approximation to the Bethe-Salpeter (BS) approach
successfully reproduces previous SU(3) WT results for the lowest-lying
s--wave negative parity baryon resonances. It also provides some
information on the dynamics of heavier ones and of the lightest d-wave
negative parity resonances, as e.g. the $\Lambda(1520)$. For charmed
baryons the scheme is consistent with heavy quark symmetry, and
our preliminary results in the strangeness-less charm $C=+1$ sector
describe the main features of the three-star $J^P=1/2^-$
$\Lambda_c(2595)$ and $J^P=3/2^-$ $\Lambda_c(2625)$ resonances. We also
find a second broad $J^P=1/2^-$ state close to the $\Lambda_c(2595)$.

\end{abstract}


\maketitle

\section{$\chi$SU(6)-BS Model}	

\begin{figure}[t]
\begin{center}
  \resizebox{75mm}{!}{\includegraphics{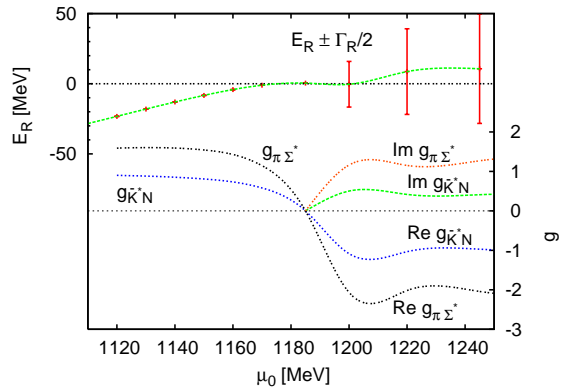}}
  \caption{{\small Results\protect\cite{l1520} for the 
$3/2^-$ $\Lambda(1520)$
   pole position ($M_R=m_\pi+M_{\Sigma^*}+E_R$,  $\Gamma_R$)
  and couplings to the $\pi\Sigma^*$ and $\bar
  K^* N$ channels as function of the renormalization scale $\mu_0$.}
  }
  \label{fig:gigj}
\end {center}
\end{figure}

A consistent SU(6) extension of the WT SU(3) chiral Lagrangian has
been recently derived in Ref.~\cite{garcia06-su6}.
The building blocks of this extension are the \{35\} (for mesons) and
\{56\} (for baryons) representations\footnote{We label the SU(N)
multiplets by their dimensionality enclosed
between curly brackets.} of SU(6). These representations can accommodate the
$0^-$ meson octet ($K$, $\pi$, $\eta$ and $\bar K$), and the
$1^-$ meson nonet ($K^*$, $\rho$, $\omega$, $\bar K^*$ and $\phi$);
and the the spin $1/2^+$ members of the nucleon octet ($N$, $\Sigma$,
$\Lambda$ and $\Xi$) and the spin $3/2^+$ members of the $\Delta$
decuplet ($\Delta$, $\Sigma^*$, $\Xi^*$ and $\Omega$), respectively. 
The scheme of Ref.~\cite{garcia06-su6}
 starts by assuming that the $s$-wave effective
meson--baryon Hamiltonian is SU(6)-spin-flavor invariant, 
and it makes use of the
underlying chiral symmetry  to determine the value of the SU(6)
irreducible matrix elements from the SU(3)-flavor WT interaction. This
is possible since the WT Lagrangian is not just SU(3) symmetric but
also chiral ($\text{SU}_L(3)\otimes\text{SU}_R(3)$)
invariant. Symbolically, 
\begin{equation}
{\cal L_{\rm WT}^{\rm SU(3)}}= {\rm Tr} \left( [M^\dagger, M][B^\dagger,
B]\right)
\end{equation}
This structure, dictated by chiral symmetry, is more suitably analyzed
in the $t$-channel. The mesons $M$ fall in the SU(3) representation $\{8\}$
which is also the adjoint representation. The commutator
$[M^\dagger,M]$ indicates a $t$-channel coupling to the $\{8\}_a$
(antisymmetric) representation, thus
\begin{equation}
{\cal L_{\rm WT}^{\rm SU(3)}}= \left[(M^\dagger\otimes M)_{\{8\}_a}\otimes
(B^\dagger\otimes B)_{\{8\}}\right]_{\{1\}}
\end{equation}
Since the \{35\} is the adjoint representation of SU(6), the unique
SU(6) extension is 
\begin{equation}
{\cal L_{\rm WT}^{\rm SU(6)}}= \left[(M^\dagger\otimes M)_{\{35\}_a}\otimes
 (B^\dagger\otimes B)_{\{35\}}\right]_{\{1\}}
\label{eq:1}
\end{equation}
The potentials, $V_{IY}^J(s)$, deduced from this
SU(6) Lagrangian\footnote{ Some
explicit SU(6) breaking effects, due to the use of physical
(experimental) hadron masses and meson decay constants, are taken into
account\cite{l1520}.} are used to solve the coupled channel Bethe Salpeter
Equation (BSE)\cite{l1520,garcia04} within the so-called on shell
renormalization scheme\cite{LK02,LK02b}, leading to unitarized
s-wave meson--baryon scattering amplitudes $T_{IY}^J$ ($I,Y,J$ are
the isospin, hypercharge and total angular momentum)
\begin{equation}
T_{IY}^J(s)=\frac{1}{1-V_{IY}^J(s)J_{IY}^J(s)}V_{IY}^J(s) \label{eq:t-matrix}
\end{equation}
where $J_{IY}^J(s)$ is a diagonal loop function in the coupled channel
space, which needs to be renormalized\cite{l1520,nieves01}. In a given
$JIY$ sector, physical resonances appear as complex poles in the
second Riemann sheet\cite{nieves01} of all matrix elements of $T(s)$
in the coupled channel space.  The pole position
determines the mass and width of each resonance, while the different
residues for each meson-baryon channel give the respective couplings
and branching ratios\cite{nieves01b}.  This model\cite{l1520,garcia07}
reproduces the essential features of previous
studies\cite{garcia04,LK02,LK02b,nieves01,nieves01b,kaiser95,kaiser95b,oset98,oset98b,oset98c,oller01}
(properties of the lowest lying $1/2^-$ and $3/2^-$ resonances) and,
in addition, it sheds some light on the role played by the vector
mesons in this context. For instance, we show in Fig.~\ref{fig:gigj} some
predictions for the d--wave $\Lambda(1520)$ resonance\cite{l1520}. We
naturally find that this resonance is placed quite close to the
$\pi\Sigma^*$ threshold and that its coupling to $\bar K^* N$ channel
is quite small. Indeed, we obtain a coupling much smaller than the
value used by phenomenological studies of the $\Lambda(1520)$
photoproduction\cite{nam05,titov05}, which brings in interesting
conclusions on the dominant reaction mechanism in these
processes\cite{l1520}.

\section{$\chi$SU(8)-BS Model} 
The extension of the above model to describe charmed baryon resonances
is straightforward. We use the \{63\} and \{120\} SU(8) representations
to accommodate mesons and baryons, respectively. In addition to those
particles considered so far, these representations also contain the
$0^-$ $D_s, D,\eta_c, {\bar D}, {\bar D}_s$ and $1^-$ $D_s^*,
D^*,J/\Psi, {\bar D}^*, {\bar D}_s^*$ mesons, and the 
$1/2^+$ $\Xi_{cc}, \Omega_{cc}, \Lambda_c,\Sigma_c,\Xi_c,\Xi'_c,\Omega_c$ and
the $3/2^+$ $\Omega^*_{ccc}, \Xi^*_{cc}, \Omega^*_{cc}, \Sigma^*_c,\Xi^*_c,
\Omega^*_c$ charmed baryons. Since
the \{63\} is the SU(8) adjoint representation, from the discussion
above, the natural extension of the WT lagrangian to the SU(8)
spin-flavor symmetry group is
\begin{equation}
{\cal L_{\rm WT}^{\rm SU(8)}}= \left[(M^\dagger\otimes M)_{\{63\}_a}\otimes
 (B^\dagger\otimes B)_{\{63\}}\right]_{\{1\}}
\end{equation}
We use the above interaction, with some explicit SU(8) breaking
effects induced by the use of experimental hadron masses and meson
decay constants, to construct the kernel of the corresponding two
body meson-baryon BSE. So far, we have only analyzed the
strangeness-less charm $C=+1$ sector, where we have naturally found
the $1/2^-$ $\Lambda_c(2595)$ (with a very small coupling to the open
$\Sigma_c\pi$ channel) and the $3/2^-$ $\Lambda_c(2625)$
resonances. These resonances have a clear parallelism with the four
star resonances $\Lambda(1405)$ and $\Lambda(1520)$, which appear in
the (charmless) strangeness $-1$ sector. Besides,  similarly to
what happens in the case of the $\Lambda(1405)$ resonance, we have
also found a second (broad) $J^P=1/2^-$ state close to the
$\Lambda_c(2595)$, but with a much larger coupling to the open $\pi
\Sigma_c$ channel.

There have been several works\cite{KL04,HL05,HL06} on $s-$ and
$d-$wave negative parity charmed baryon resonances. In all these works
the zero-range $t$-channel exchange of vector mesons is identified as
the driving force for the $s-$wave scattering of pseudo-scalar mesons
off the baryon ground states.  A serious limitation of this
$t$-channel vector meson-exchange (TVME) model is that while the
pseudo-scalar mesons $D$ and $D_s$ are included in the coupled-channel
dynamics, their vector partners $D^*$ and $D_s^*$ are completely
ignored.  This is not justified since QCD acquires a new spin-flavor
symmetry\cite{IW89}, Heavy Quark Symmetry (HQS), when the quark masses
are much larger than the typical confinement scale, $\Lambda_{\rm
QCD}$. HQS predicts that all type of spin interactions vanish for
infinitely massive quarks. Neglecting corrections of the order
$\Lambda_{\rm QCD}/m_c$, the $D$ and $D^*$ or $D_s$ and $D^*_s$ mesons
form a multiplet of degenerate hadrons\cite{IW89}.  For finite charm
quark mass, the  $D$ and the  $D^*$ meson masses differ in
just about one pion mass, even less for the strange charmed mesons,
and thus it is reasonable to expect that the coupling
$DN-D^*N$ might play an important role.

Our scheme naturally respects HQS, which justifies, at least in the
charm sector, the assumption that the quark interactions are spin
independent. The inclusion of $D^*$ degrees of
freedom in the coupled channel formalism changes the dynamics. For
instance, we find that $D^*N$ component, absent in the TVME model, of
the $\Lambda_c(2595)$ is sizable and much larger than the
 $DN$ and the $\Sigma_c\pi$ ones. As a consequence, we
predict a totally different dynamical picture for this resonance.

\section*{Acknowledgments}

This work has been supported by the Spanish Consolider-Ingenio 2010
Programme CPAN (CSD2007-00042), Junta de Andaluc\'\i a grant
FQM0225, MEC grants FIS2005--00810, FIS2005--03142 and
from the EU Human Resources and Mobility Activity, FLAVIAnet, contract
number MRTN--CT--2006--035482.


\begin{thebibliography}{0}

\bibitem{garcia06-su6} C. Garc\'\i a-Recio, J. Nieves and L.L. Salcedo, 
                              {\it Phys. Rev. D} {\bf 74} 034025 (2006).

\bibitem{l1520} H. Toki, C. Garc\'{\i}a-Recio, and J. Nieves, 
{\it Phys. Rev. D} {\bf 77} 034001 (2008).


\bibitem{garcia04} C. Garc\'{\i}a-Recio, M.F.M Lutz and J. Nieves, 
{\it Phys. Lett. B} {\bf 582} 49 (2004).

\bibitem{LK02} M. F. M. Lutz and E. E. Kolomeitsev, {\it Nucl. Phys. A} 
{\bf 700} 193 (2002).

\bibitem{LK02b} E.E. Kolomeitsev and M.F.M. Lutz, {\it
  Phys. Lett. B} {\bf 585} 243 (2004).


\bibitem{nieves01} J. Nieves and E. Ruiz-Arriola, {\it Phys. Rev. D} {\bf 64}
   116008 (2001).



\bibitem{nieves01b} C. Garc\'{\i}a-Recio et al.,  {\it Phys. Rev. D } {\bf 67}  076009 (2003).


\bibitem{garcia07} C. Garc\'{\i}a-Recio, J. Nieves and L.L. Salcedo, 
 {\it Eur. Phys. J. A} {\bf 31} 499 (2007).



\bibitem{kaiser95} N. Kaiser, P.B. Siegel and W. Weise, {\it
  Nucl. Phys. A} {\bf 594} 325 (1995).

\bibitem{kaiser95b} N. Kaiser, P.B. Siegel and W. Weise, {\it Phys. Lett. B} 
{\bf 362} 23  (1995).

\bibitem{oset98} E. Oset and A. Ramos, {\it Nucl. Phys. A  } {\bf 635} 99
  (1998).

\bibitem{oset98b} D. Jido, et al., 
{\it Nucl. Phys. A} {\bf 725} 181 (2003).

\bibitem{oset98c} S. Sarkar, E. Oset and
M.J. Vicente--Vacas, {\it Nucl. Phys. A} {\bf 750} 294 (2005).

\bibitem{oller01} J. Oller and U. Meissner, {\it Phys. Lett. B} 
{\bf 500} 263 (2001).


\bibitem{nam05} S.I. Nam, A. Hosaka and H.C. Kim, {\it  Phys. Rev. D} {\bf 71}
 114012  (2005).


\bibitem{titov05} A.I. Titov, B. K\"ampfer, S. Date and Y. Ohashi, 
{\it Phys. Rev. C} {\bf 72}  035206 (2005).

\bibitem{KL04} M.F.M. Lutz and E.E. Kolomeitsev, {\it Nucl. Phys. A} {\bf 730}
  110 (2004).

\bibitem{HL05} J. Hofmann and M.F.M. Lutz, {\it Nucl. Phys. A} {\bf 763}
 90 (2005).

\bibitem{HL06} J. Hofmann and M.F.M. Lutz, {\it Nucl. Phys. A} {\bf 776}
  17 (2006).

\bibitem{IW89} N. Isgur and M.B. Wise, Phys. Lett. {\bf B232} (1989) 113.

\end{thebibliography}
\end{document}